\documentclass[nohyper]{JHEP}
\usepackage{amsmath,amssymb,epsfig,eucal}

\psfull

\newcommand{\ga}{\gamma}
\newcommand{\gat}{{\tilde \gamma}}

\newcommand{\om}{\omega}
\newcommand{\as}{\alpha_{\mathrm{s}}}
\newcommand{\asb}{{\bar \alpha}_{\mathrm{s}}}
\def\cO#1{{\cal O}\left(#1\right)}
\renewcommand{\d}{\mathrm{d}}
\newcommand{\cF}{{\cal F}}
\newcommand{\cG}{{\cal G}}
\newcommand{\cN}{{\cal N}}

\newcommand{\vk}{{\vec k}}
\newcommand{\vq}{{\vec q}}
\newcommand{\Qbar}{{\bar Q}}
\newcommand{\eff}{\mathrm{eff}}
\newcommand{\ie}{\textit{i.e.}\ }

\newcommand{\half}{\mbox{$\frac{1}{2}$}}
\newcommand{\cmF}{{\CMcal F}}
\newcommand{\cmI}{{\CMcal I}}
\newcommand{\cmN}{{\CMcal N}}
\newcommand{\cmD}{{\CMcal D}}

\newcommand{\tbar}{{\bar t}}
\newcommand{\omp}{{\omega_{\mathbb{P}}}}

\title{On factorisation at small $\pmb x$\thanks{Work
    supported by E.U. QCDNET contract FMRX-CT98-0194 and by the
    Fondazione A.~della Riccia.}}
\author{Marcello Ciafaloni$^1$, Dimitri Colferai$^{1,2}$ and Gavin
    P. Salam$^3$\\ 
  $^1$ Universit\`a di Firenze and INFN, Sezione di Firenze,
  Firenze, Italy.\\
  $^2$ II. Institut f\"ur theoretische Physik, Hamburg, Germany.\\
  $^3$ TH Division, CERN, 1201 Geneva 23, CH.}

\abstract{ We investigate factorisation at small $x$ using a variety
  of analytical and numerical techniques. Previous results on
  factorisation in collinear models are generalised to the case of the
  full BFKL equation, and illustrated in the example of a collinear
  model which includes higher twist terms.  Unlike the simplest
  collinear model, the BFKL equation leads to effective anomalous
  dimensions containing higher-twist pieces which grow as a
  (non-perturbative) power at small $x$. While these pieces dominate
  the effective splitting function at very small $x$ they do not lead
  to a break-down of factorisation insofar as their effect on the
  predicted scaling violations remains strongly suppressed.  }


\preprint{
  CERN--TH/2000--209\\
  DFF364/9/2000\\
  DESY 00--117\\
  hep-ph/0007240 \\
  July 2000
  }

\begin{document}

\section{Introduction}
\label{sec:intro}

The standard approach to calculating any quantity in Deep Inelastic
Scattering (DIS) or Drell-Yan type processes is to assume
factorisation, based on the operator product expansion. This allows us
to consider parton distributions which evolve with scale as 
\begin{equation}
  \label{eq:fact}\
  \frac{\partial g_{\om}(Q^2)}{\partial \ln Q^2} = \ga_\om(Q^2)
  g_{\om}(Q^2) + \cO{\frac{1}{Q^2}}, 
\end{equation}
where $\om$ is the Mellin moment of the parton distribution. Cross
sections are then expressed in terms of the parton distribution at
scale $Q^2$, multiplied by a coefficient function, which is itself a
function of only $\om$ and $\as(Q^2)$, plus corrections of order
$1/Q^{2n}$.

A few years ago however, it was suggested in \cite{Mueller} that at
small $x$ (small $\om$) diffusion might place a limit on
factorisation: namely that beyond a certain value of $x$ factorisation
would break down. The argument was the following: at moderate values
of $x$ parton evolution in $Q^2$ can be associated with a chain of
emissions ordered in $Q^2$. Corrections to factorisation can arise
from chains which evolve from some starting scale $Q_{\textrm{start}}^2$
up to a scale $Q^2$, evolve down to a non-perturbative scale
$\Lambda^2$ and then evolve back up to scale $Q^2$. At moderate $x$,
such contributions are suppressed by a power of $\Lambda^2/Q^2$. At
small $x$ however, this process of going up and down in scale becomes
an essential element of the evolution, and is no longer suppressed by
powers of $Q^2$: so it would seemed that factorisation might break down
at the values of $x$ such that this diffusion enters into the
non-perturbative region. More recently this has been used as an
argument for parameterising and fitting the small-$x$ behaviour of
splitting functions, rather than trying to calculate them from first
principles \cite{ABF}.

In \cite{CCS1999a} plausibility arguments were presented suggesting
that there would be no such breakdown. Physically speaking, perhaps
the simplest explanation is to say that evolution that, while going
down in $x$ drops below one's starting scale $Q_{\textrm{start}}^2$,
contributes to the small-$x$ part of the gluon distribution at the
starting scale. Therefore it should not be included in the anomalous
dimension in order to avoid counting it twice, so the anomalous
dimension contains only evolution which remains above the starting
scale, ensuring that it is perturbative.

However this explanation, and the arguments in \cite{CCS1999a}, are
not entirely satisfactory because they are not able to estimate the
size or functional dependence on $x$ of any violations of
factorisation.

In \cite{CCS1999b}, a model was studied which was qualitatively
similar to BFKL in that it had diffusion, and which correctly
reproduced collinear evolution at all orders. It was possible to show
that this model leads to \emph{exact factorisation}, namely that there
were no $1/Q^2$ corrections at all. Thus diffusion on its own is not
incompatible with factorisation.

This collinear model had two particularities: the branching kernel,
when converted to $\gamma$ space (the Mellin transform variable
conjugate to transverse momentum) contained poles only at $\ga=0$ and
$\ga=1$, \ie the leading collinear and anti-collinear poles. The BFKL
equation instead has poles at all integer values of $\ga$. The second
particularity was related to the scale of $\as$: in the collinear
model the scale was chosen to be $\max(k^2,{k'}^2)$, where $k$ and
$k'$ are the exchanged transverse momenta before and after the
branching. In the BFKL equation, the next-to-leading corrections seem
to indicate that the correct scale is rather $q=|{\vec k} - {\vec
  k}'|$ \cite{NLLscale} which has the property that it can go to zero
even when both $k$ and $k'$ are large.

Thus there is a need for a study of factorisation within the BFKL
equation. The complications mentioned above make this quite difficult
to do analytically. We therefore adopt a two-pronged approach. In
section \ref{sec:4pole} we consider an extension of the
collinear model with additional poles at $\ga=-1$ and $\ga=2$. Some of 
its gross features can be deduced analytically and can be expected to
carry over to the full BFKL equation. 

In section \ref{sec:bfkleqn} we then present a simple numerical method
which for the first time allows the extraction of the exact (effective
--- see below) anomalous dimensions in the full leading logarithmic
BFKL equation, with all-order running coupling corrections, allowing a
detailed study of factorisation not limited by the simplifying
assumptions of any given particular model.


\section{Understanding factorisation}
\label{sec:understanding}

We shall start off by discussing factorisation for the non-integrated
gluon distribution, or equivalently for the gluon Green's function.
Conceptually it it somewhat simpler --- the generalisation to the
integrated gluon distribution is then relatively straightforward.

We shall discuss factorisation by considering a number of models. We
study a gluon Green's function $G_\om(t,t_0)$, $t = \ln Q^2$, in
analogy with the unintegrated gluon distribution in the proton. and
consider as the non-perturbative aspects of the problem the value of
$t_0$ (the lower hard scale of the process) and the regularisation 
of the running coupling in the infra-red.
The latter will be represented by some scale $\tbar$ at which the
coupling is cut off (we could equally have chosen to freeze the
coupling at that scale). The small-$x$ properties of the Green's
function will depend on both $t_0$ and $\tbar$. More specifically the
position of the leading pole in the $\om$-plane will depend on the
value of $\tbar$ and its normalisation on both $\tbar$ and $t_0$.

If factorisation holds then the (non-integrated) effective anomalous
dimension defined by
\begin{equation}
  \label{eq:gatDef}
  \gat_\om(t) = \frac{1}{G_\om(t,t_0)} \frac{\partial G_\om(t,t_0)}{
      \partial t}\,,
\end{equation}
should be independent of both $\tbar$ and $t_0$ at least to within
higher-twist terms suppressed at least as $e^{-t}$.

This is not quite a sufficient condition: indeed if the anomalous
dimension contains higher-twist pieces which grow as a sufficiently
large power of $x$ then these could dominate the scaling violations
making them impossible to predict. Since the Pomeron singularity will
affect $G_\om$ anyway, yielding a large power behaviour of this kind,
the problem arises of understanding whether approximate factorisation
is still preserved in the small-$x$ region.

Since we are interested in pieces enhanced at small-$x$ we need to
understand the singularity structure of the anomalous dimension.  
There are two possible origins for singularities of
$\gat_\om$. One possible origin is that at some $\om$, $\partial
G_\om/G_\om$ contains a non-factorisable singularity, 
the other is for $G_\om$ to be zero while $\partial
G_\om$ is finite.

Now, the simplest collinear model of \cite{CCS1999b} provided us with
a mechanism by which the Pomeron singularity in $G_\om$ can be
consistent with exact factorisation: the leading singularity of
$\gat_\om$ was found to come from a $t$-dependent \emph{zero} of
$G_\om$, leaving the Pomeron factorised away. We shall see in the
following that this basic mechanism is still at work in collinear
models with higher twist terms \cite{cc} and in the BFKL equation
itself, thus suppressing the non-factorisable singularities.

\subsection{Recalling the structure of the collinear model}
\label{sec:coll}

The collinear model of ref.~\cite{CCS1999b} gave a powerful handle for
the study of anomalous dimensions. It was described by an equation of
the form
\begin{equation}
  \label{eq:2poleint}
  \om G_\om(t,t_0) = \delta(t - t_0) +
  \int dt'  K_2(t,t')\, G_\om(t',t_0),
\end{equation}
whose kernel is
\begin{equation}
  \label{eq:2poleK}
  K_2(t,t') = \asb(t) \Theta(t-t')+ 
  \asb(t') \Theta(t'-t)  e^{-(t'-t)}.
\end{equation}
In $\gamma$-Mellin transform space with respect to $Q^2 = \exp(t)$,
the leading order part of the kernel gives the following
characteristic function:
\begin{equation}
  \label{eq:chi2pole}
  \chi^{\mathrm{2-pole}}(\ga) = \frac1\ga + \frac1{1-\ga}\,,
\end{equation}
for which reason it is also referred to as a $2$-pole model. It has
the convenient property that it can be expressed in terms of a
second order differential equation, whose solutions have the following
factorised form:
\begin{equation}
  \label{eq:G2pole}
  G_\om(t,t_0) = F^L_\om(t_0) F^R_\om(t), \qquad t > t_0\,,
\end{equation}
where $F_\om^L$ and $F_\om^R$ are the linearly independent
solutions of the homogeneous equation (eq.~\eqref{eq:2poleint} without
the $\delta$-function term), that are regular to the left (negative
$t_0$) and to the right (positive $t$) respectively.

The fact that $F_\om^R(t)$ is regular at large $t$ means that it is
independent of $\tbar$ for $t> \tbar$. It is free of singularities in
$\om$-space, as illustrated in figure~\ref{fig:airyplots}.  All
non-perturbative dependence, in particular the poles governing the
high-energy behaviour of $G$, is contained in $F_\om^L$, also
illustrated in figure~\ref{fig:airyplots}.

One sees from these plots that zeroes of $G_\om$ can arise from both
$F_\om^L$ and $F_\om^R$. \footnote{Note however that $F_\om^L$ does
  not always have zeroes: their presence depends on a variety of
  factors such as the relative sign of successive divergences, which
  are determined by the non-perturbative parameters. Only in the
  situation of $t_0 \gg \tbar$ is at least one zero guaranteed, just
  to the left of the leading divergence.}  However since the
$t$-dependence lies entirely in $F_\om^R$, the $t_0$-dependent,
non-factorisable zeroes of $F_\om^L$ lead to
zeroes of both $G_\om$ and of its $t$-derivative.  Therefore they do
not lead to divergences of the anomalous dimension.  Only the $t$-dependent
zeroes of
$F_\om^R$ lead to poles in $\ga(\om)$, $\partial_t F_\om^R(t)$ not
usually being zero when $F_\om^R(t)$ is zero.

This fact has the consequence mentioned before: the Pomeron singularity,
present in $F^L_\om(t_0)$, does not occur in the anomalous dimension,
which is singular at $\om_c(t)$,
the leading zero of $F^R_\om$ (while we refer to
the leading zero of $F^L_\om(t_0)$ as $\om_0(t_0)$).

\FIGURE{
    \epsfig{file=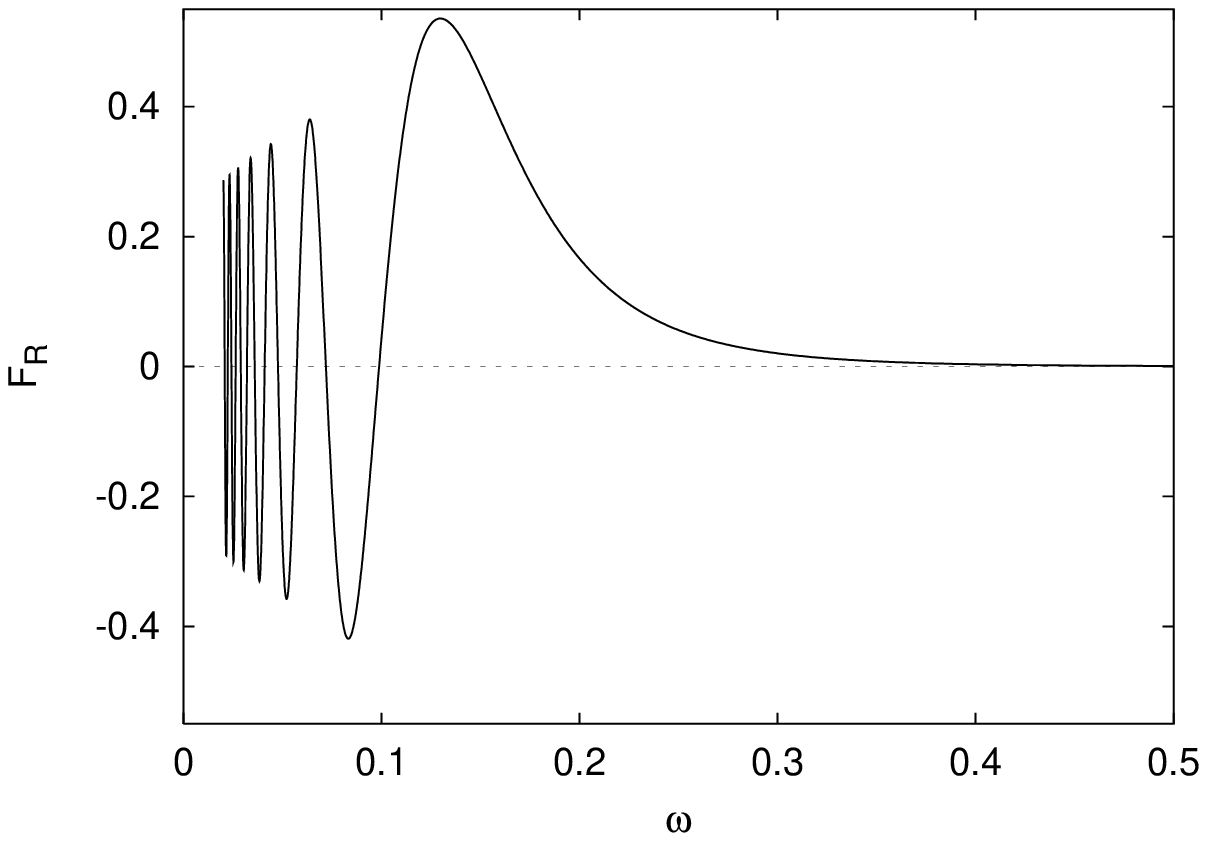, width=0.71\textwidth}\\
    \mbox{\;\;\;}\epsfig{file=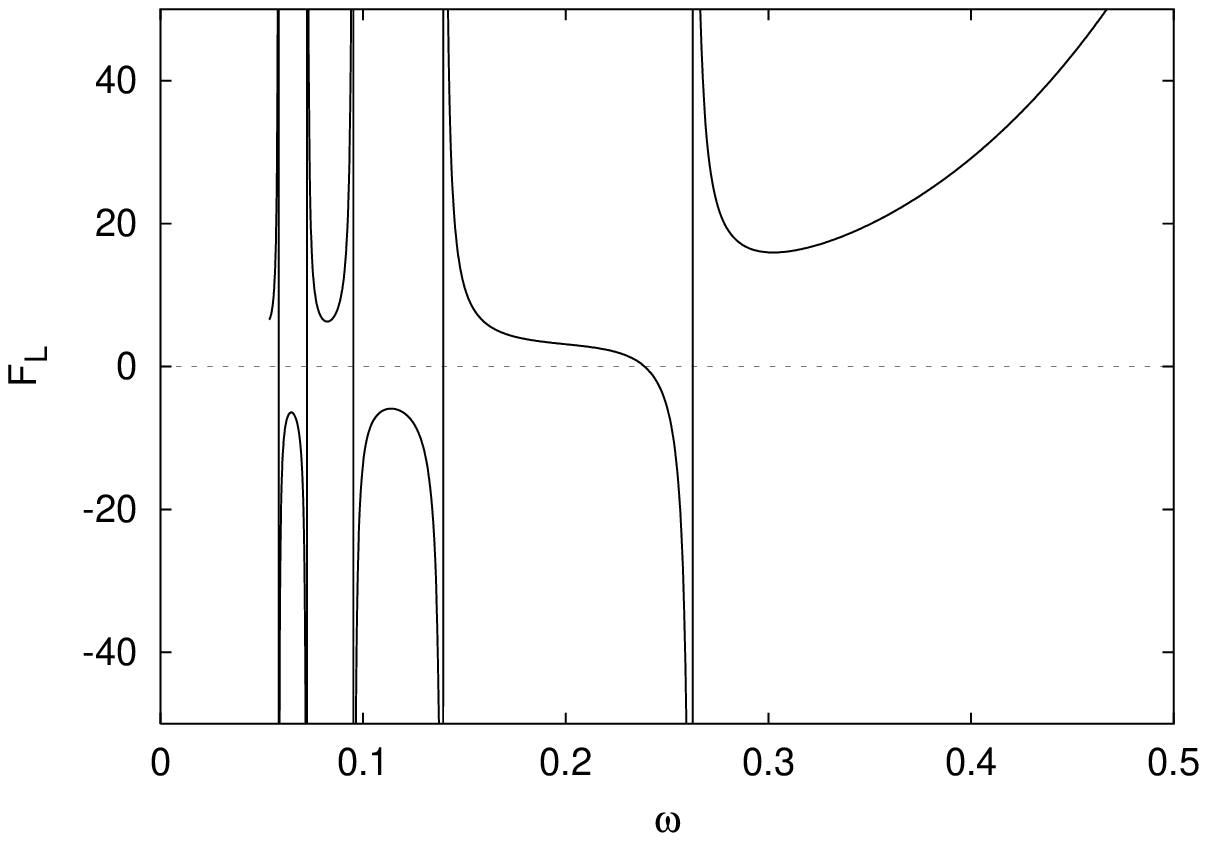,width=0.7\textwidth}\\
    \caption{Illustration of the kind of the $\om$-dependence that
      arises in the right- and left-regular solutions of models based
      on second-order differential equations such as the two-pole
      collinear model. (The small-$\om$ parts of the curves have been
      left out in order to improve the overall clarity of the plots).
      }
    \label{fig:airyplots}
}

\section{The 4-pole collinear model}
\label{sec:4pole}

One of the main differences between the BFKL equation and the
collinear model just-discussed lies in the presence in the BFKL
equation of poles of $\chi(\ga)$ at all integer values of $\ga$. The
poles beyond $\ga=0$ and $\ga=1$ can give rise to higher-twist
effects, in which we are particularly interested in this article. 

A useful stepping stone to the full BFKL equation is an equation
containing the first subleading higher-twist parts, namely poles at
$\ga=-1$ and $\ga=2$. The equation for the Green's function is
\begin{equation}
  \label{eq:4poleint}
  \om G_\om(t,t_0) = \delta(t - t_0) +
  \int dt'  K_4(t,t')\, G_\om(t',t_0),
\end{equation}
where the kernel is
\begin{multline}
  \label{eq:4poleK}
  K_4(t,t') = \asb(t) \Theta(t-t') \left(1 + e^{-(t-t')}\right) + 
  \asb(t') \Theta(t'-t) \left(e^{-(t'-t)} + e^{-2(t'-t)}\right)\\
  - \frac43 \asb(t) \delta(t-t')\,.
\end{multline}
The $\delta$-function term is included so that the value of
$\chi_0(1/2)$ be the same as for the 2-pole collinear model:
\begin{equation}
  \label{eq:chi4pole}
  \chi^{\mathrm{4-pole}}(\ga) = \frac1{1+\ga} + \frac1\ga +
   \frac1{1-\ga} + \frac1{2-\ga} - \frac43.
\end{equation}
The physics content of this formulation is quite similar to the
leading and first-subleading collinear and anti-collinear parts of the
BFKL equation. A slight difference exists in that the
higher-order higher-twist pieces of the BFKL equation also contain a
non-local dependence on $\as$. Their inclusion would considerably
increase the difficulty of the analytic treatment.

This model as it stands can be expressed as a fourth-order
differential equation (or two coupled second order differential
equations, etc.), using the same approach as in \cite{CCS1999b}, as
discussed in detail in the Appendix. It is thus similar to the
diffusion models with higher twist terms \cite{cc}. Without having to
write down the full equations in detail, we describe here some
important properties of its solutions.

There are four linearly-independent solutions of the homogeneous
equation. We denote them by $F^{R,0}$, $F^{R,1}$, the leading and
sub-leading twist right-regular solutions, which for large $t$ go as a
constant and $e^{-t}$ respectively; and by $F^{L,a}$, $F^{L,b}$ for
the two left-regular solutions. For $t> t_0$ the Green's function 
has the expression
\begin{equation}
  \label{eq:4poleG}
  G_\om(t,t_0) =  F_\om^{L,a}(t_0) F_\om^{R,0}(t) + F_\om^{L,b}(t_0)
  F_\om^{R,1}(t)\,. 
\end{equation}
as expected from \cite{cc}, and proved in the Appendix. There we also
show how to determine the $F^{L}$'s so as to satisfy the boundary
conditions at $t=t_0$. The solution for $t<t_0$ can then also be
obtained, by a simple exchange of variables.  As in the 2-pole model,
the right-regular solutions are independent of the non-perturbative
parameters of the problem. The Pomeron singularity and the
non-perturbative dependence all enter into the left-regular solutions.

It is convenient to divide the solution into its leading and
higher-twist parts, $G_\om^0$ and $G_\om^1$, corresponding to the
first and second terms of \eqref{eq:4poleG} respectively.  For large
$t\gg t_0, \tbar$, we have that $G^1_\om$ is strongly suppressed (by a
relative amount $\sim e^{-t}$) compared to $G^0_\om$; $G^0_\om$ should
be qualitatively similar to the solution of the 2-pole model, having
zeroes of perturbative and non-perturbative origin. We know
therefore, from the start, that the violation of factorisation is
uniformly of higher twist, just because of the additive nature of
\eqref{eq:4poleG}.

On the other hand, one can look at the same problem from the standpoint 
of the singularities of the anomalous dimension. Let us concentrate on 
the case with a discrete spectrum (running coupling with cutoff), in which
we have to look for zeroes of $G_\om$.
The fact that $G_\om^1$ is strongly suppressed means that the
distribution of zeroes of $G_\om$ is determined essentially by the
positions of the zeroes of $G_\om^0$. But in general where $G_\om^0$
is zero, $G_\om^1$ will be non-zero, causing the zeroes of $G_\om$ to
be slightly shifted compared to the zeroes of $G_\om^0$. The size of
this shift will be related to the relative sizes of $G_\om^1$ and
$G_\om^0$, \ie it will be of the order of of $e^{-t}$.

The effect of the shift will be different according to whether the
zero is of perturbative or non-perturbative origin. In the case of the
leading zero of perturbative origin, then the leading perturbative
pole of the anomalous dimension is shifted, and its normalisation 
changes --- both the effects are of order $e^{-t}$ (the relation
between relative changes to the normalisation and the position of the
divergence depends on the nature of the $t$ and $\om$-dependence of
$G^1$, and so is difficult to predict).


The consequences of the rightmost (NP) zero $\om_0(t_0)$ being shifted
are somewhat more interesting, and possibly dangerous for
factorisation. We recall that in the two-pole model, the
non-perturbative zeroes did not lead to divergences of the anomalous
dimension because the zero was in $F^L$ and so the derivative
$\partial_t G(t,t_0) = F^L(t_0) \partial_t F^R(t)$ was also zero when
$G$ was zero. This is no longer true when higher twist terms are
present, and a singularity around $\om_0$ is expected. The exact value
of $\om_0$, and indeed even the existence of this zero, depend
rather subtly on the values of $\tbar$ and $t_0$. However, for $t_0$
significantly larger than $\tbar$ one expects that the zero
exists, and is driven by the Pomeron term in $F^L(t_0)$, so that 
$\omp-\om_0 \sim e^{-t_0}$ is rather small and $\om_0$ may be leading.


To see in detail what happens in the 4-pole model if $\om_0$
is leading, we approximate
$G^0_{\om}(t,t_0)$ around its zero by $\cN_0 (\om - \om_0)
F^{R,0}_{\om_0}(t)$, where $\cN_0 = \partial_\om
F_\om^{L,a}(t_0)|_{\om=\om_0}$, and $G^1(t,t_0)$ by $\cN_1
F^{R,1}_{\om_0}(t)$. It is useful to define partial anomalous
dimensions separately for $F^{R,0}_{\om_0}(t)$ and
$F^{R,1}_{\om_0}(t)$:
\begin{align}
  \ga^0_\om(t) &= \frac{\partial_t
  F^{R,0}_{\om}(t)}{F^{R,0}_{\om}(t)} = \frac{\asb}{\om} +
  \ldots,
  \\
  \ga^1_\om(t) &= \frac{\partial_t
  F^{R,1}_{\om}(t)}{F^{R,1}_{\om}(t)} = -1 + \frac{\asb}{\om} +
  \ldots,
\end{align}
where we have neglected terms of order $(\asb/\om)^2$ in the expansion
of the anomalous dimensions.  Thus $F^{R,1}_{\om_0}(t)$ is suppressed
compared to $F^{R,0}_{\om_0}(t)$ by an amount
\begin{equation}
  \frac{F^{R,1}_{\om_0}(t)}{F^{R,0}_{\om_0}(t)} 
  \propto 
  \exp\left[ 
  \int^t dt' \left(\ga^1_{\om_0}(t) \,-\, \ga^0_{\om_0}(t)\right) \right]
  \sim e^{-t \,+\, \cO{\as(t)/\om_0^2}}\,.
\end{equation}
The rightmost zero of $G_\om(t,t_0)$, $\om_0'$, is therefore shifted
compared to that of $F^{R,1}_{\om}(t)$ by an amount
\begin{equation}
  \label{eq:newzero}
  \om_0' - \om_0 \simeq  - \frac{\cN_1 F^{R,1}_{\om_0}(t)}{\cN_0
  F^{R,0}_{\om_0}(t)} \sim e^{-t \,+\, \cO{\as/\om_0^2}}\,. 
\end{equation}
At $\om_0'$ the $t$ derivative of $G_\om$ is non-zero, leading to a
pole of the anomalous dimension, whose residue $R(\om_0')$ is
\begin{equation}
  \label{eq:residueom0}
  R(\om_0') = (\ga^1_{\om_0}(t) - \ga^0_{\om_0}(t)) \frac{\cN_1
  F^{R,1}_{\om_0}(t)}{\cN_0 
  F^{R,0}_{\om_0}(t)} = \left(1 + \cO{\as^2/\om^2} \right) (\om_0' -
  \om_0) 
\end{equation}
So we have that the residue of the leading NP pole is the same as the
shift of the zero and \emph{both are higher twist}. In the case of the BFKL
equation (section~\ref{sec:bfkleqn}) one can test to see if this
remains true, in order to verify that the same mechanisms are at work
there as in the 4-pole model.\footnote{We note that for the BFKL
  equation including next-to-leading corrections, the relation between
  $\om_0'-\om_0$ and the residue of the pole can be modified by pieces
  of relative order $\as$.}


Since $\om_0$, which may approach $\omp$, is likely to be to the right
of the perturbative pole $\om_c$, at small $x$ this higher-twist
non-perturbative contribution will dominate the anomalous dimension.
At first sight this might seem to have worrying implications for the
prediction of small-$x$ scaling violations. But in the present model
we know that this is not the case, because of the validity of
\eqref{eq:4poleG}. From the point of view of $t$-evolution we are
saved by the fact that at small $x$, in the convolution of the
splitting function with the non-perturbative input distribution (which
grows as $\omp > \om_0$), the higher-twist part of the effective
splitting function gives a contribution of order
\begin{equation*}
  \frac{e^{-t}}{\omp - \om_0}
\end{equation*}
while the perturbative contribution is of order
\begin{equation*}
  \frac{\as(t)}{\omp}\,,
\end{equation*}
both growing as $x^{-\omp}$. We thus recover in the latter 
contribution the Pomeron
part of the leading twist term in \eqref{eq:4poleG}, that we know to 
be factorised.
Thus there is always a value of $Q^2$ such that the higher-twist
corrections can be ignored at all $x$.

A small point worth bearing in mind is that our analysis so far has
always been for the anomalous dimensions related to unintegrated gluon
distributions. In practice one is more interested in the anomalous
dimensions of the integrated distributions. It turns out that their
properties are very similar: this is because the $n$-pole models can be
expressed in terms of $n$ coupled linear differential equations, and
the unintegrated and integrated gluon distributions are simply
different linear combinations of the components of the equations.

\section{The BFKL equation}
\label{sec:bfkleqn}

We shall study the leading-order BFKL equation including a running
coupling, 
\begin{multline}
  \label{eq:bfklint}
  \cG(x,k,Q_0) = \delta(k^2/Q_0^2 - 1) +\int_x^1
  \frac{\d z}{z}
  \int \frac{\d^2 \vq}{\pi q^2}\, \asb(q^2) \\
  \times
  \left[\frac{k^2}{|\vk - \vq|^2}\cG(x,|\vk - \vq |,Q_0) -
    \Theta(k-q)\cG(x,k,Q_0)\right]. 
\end{multline}
For notational convenience we have switched to using transverse
momenta $k,Q_0$, rather than the logs of their squares $t,t_0$.  The
choice of the emitted transverse momentum as the scale for the running
coupling, $\as(q^2)$, is suggested by the form of the NLO corrections
to the kernel \cite{NLLscale}.  A normal DGLAP gluon distribution is
expressed in terms of $\cG(x,k,Q_0)$ via $k$-factorisation:
\begin{equation}
  \label{eq:kfact}
  xg(x,Q^2) = \int^{Q^2} \frac{\d^2k}{\pi k^2} \, \cG(x,k,Q_0)\,.
\end{equation}
In practice rather than solving the integral equation
\eqref{eq:bfklint}, it is easier to solve the related differential
equation
\begin{multline}
  \label{eq:bfkldif}
  \frac{\partial \cG(x,k,Q_0)}{\partial \ln 1/x} = 
  \int \frac{\d^2 \vq}{\pi q^2}\, \asb(q^2)
  \left[\frac{k^2}{|\vk - \vq|^2} \cG(x,|\vk - \vq |,Q_0) 
    - \Theta(k-q)\cG(x,k,Q_0)\right],
\end{multline}
with initial condition $\cG(1,k,Q_0) = \delta(k^2/Q_0^2-1)$.

\subsection{The extraction of anomalous dimensions}
Naively to obtain the effective splitting function, one would determine
$\cG$ and then solve for the function $P_{gg,\eff}(z)$ such that
\begin{equation}
  \label{eq:Px}
 x \partial_{\ln Q^2} g(x,Q^2) =x \int_x^1 \frac{dz}{z} P_{gg,\eff}(z)
 g(x/z,Q^2)\,. 
\end{equation}
However such an approach turns out to be subject to considerable
numerical instabilities. The reason is that any method of solution for
$\cG$ introduces small errors (typically of the relative order of
$10^{-2}$--$10^{-3}$). When carrying out the deconvolution it generally
turns out that $P(z)$ for small $z$ contributes only a small amount to
the scaling violations ($xg(x,Q^2)$ grows as $x^{-\omp}$ whereas
perturbatively, $P(z)$ grows as $x^{-\om_c}$ and $\om_c \ll \omp$).
When $x$ is such that the small error on $\cG(x)$ is of the same order
as the contribution to the scaling violations from $P(x)$, then we no
longer have a handle on the splitting function.

A solution is to choose an inhomogeneous term such that $g(x,Q^2)$ is
independent of $x$. Then, for a given $x$, the convolution
\eqref{eq:Px} is dominated by small $z$'s and small errors on $g(x)$
are no longer amplified when translated to $P(x)$. We introduce
$\cF(x,k)$ as being the unintegrated gluon distribution which,
integrated, gives $xg(x,Q^2) = 1$. It satisfies the equation
\begin{multline}
  \label{eq:cFdif}
  \frac{\partial \cF(x,k)}{\partial \ln 1/x} = f(x,k)\delta(k^2/Q_0^2-1) +
  \int \frac{\d^2 \vq}{\pi q^2}\, \asb(q^2) \\ \times
  \left[\frac{k^2}{|\vk - \vq|^2} \cF(x,|\vk - \vq |) -
    \Theta(k-q)\cF(x,k)\right], 
\end{multline}
where $\cF(1,k) = \delta(k^2/Q_0^2-1)$ and $f(x)$ is given
implicitly by
\begin{equation}
  \label{eq:fimplicit}
  f(x) = 
  -  \int^{Q^2}  \frac{\d^2k}{\pi k^2}
  \int \frac{\d^2 \vq}{\pi q^2}\, \asb(q^2)  
   \left[\frac{k^2}{|\vk - \vq|^2} \cF(x,|\vk - \vq |) -
     \Theta(k-q)\cF(x,k)\right] . 
\end{equation}
It is trivial to verify that this leads to $xg(x,Q^2) = 1$.  The form
of $f(x)$ depends on the $Q^2$ value at which we intend to consider
the splitting function and on the initial scale $Q_0$. Using
\eqref{eq:Px} it is now simple to obtain the effective splitting
function:
\begin{equation}
  \label{eq:splitfn:extrct}
  xP_{gg,\eff}(x) = \frac{\partial \cF(x,Q)}{\partial \ln 1/x}.
\end{equation}
This turns out to be numerically stable, at least until $xP_{gg,\eff}$
becomes comparable to the inverse of the machine precision.

Equation \eqref{eq:cFdif} is solved by discretising $\cF(x,k)$
uniformally in $\ln k$ space, and then applying standard Runge-Kutta
techniques for solving the resulting matrix differential equation.
This method has the advantage over other potentially faster methods,
such as a representation with a basis of Chebyshev polynomials
\cite{Lund,Durham}, that it quite easily accommodates the large
variations that arise in the value of $\cF(x,k)$ (and errors are at
worst of the relative order of the discretisation interval).

For $\asb$ we take the asymptotic formula, \ie
\begin{equation}
  \label{eq:as}
  \asb(q^2) = \frac{\Theta(q^2 - \Qbar^2)}{b \ln
  q^2/\Lambda^2},
\end{equation}
where $b = 11/12$ (we work with zero flavours). The cutoff at small
momenta corresponds to the regularisation prescription used in the
previous section for the collinear model.


\subsection{Results}
\label{sec:results}

This section has two aims. Firstly to demonstrate that for the BFKL
equation the splitting function is a truly perturbative quantity, and
that any non-perturbative dependence is higher-twist. And secondly, to 
show that our understanding of higher-twist effects as obtained from
the 4-pole model, carries over to the BFKL equation.

Let us start by examining some concrete examples of effective
splitting functions. Figure~\ref{fig:astfixed} illustrates the
effective splitting function as a function of $x$ in three situations,
all with the same value of $t$, but different sets (a, b and c) of
non-perturbative parameters, $\tbar$ and $t_0$. Going down in $x$ from
$x=1$, one sees that initially the three splitting functions are
almost identical (the inset with the larger scale reveals small
differences between them).  For moderately small $x$ the splitting
function actually decreases (this phenomenon was recently observed
also by \cite{Thorne}), and then starts to grow as $x^{-\om_c}$. The
late onset of the power growth is related to the fact that in
$\om$-space the PT pole of the anomalous dimension has a residue of
order $\asb^2$.

At a certain point, two of the curves (b and c) change sign (since we
use a logarithmic scale and plot the absolute value of the splitting
function, the change of sign appears as a downward cusp) and start to
grow with a much larger power ($\om_0'$). This is the non-perturbative
higher-twist component of the splitting function discussed in the
section~\ref{sec:4pole}. As can be seen the exact value of $\om_0'$
depends on the non-perturbative parameters. Furthermore there are
situations (curve a) in which there is no NP power growth at all,
corresponding to the absence of NP zeroes in the Green's function.

\FIGURE{
    \epsfig{file=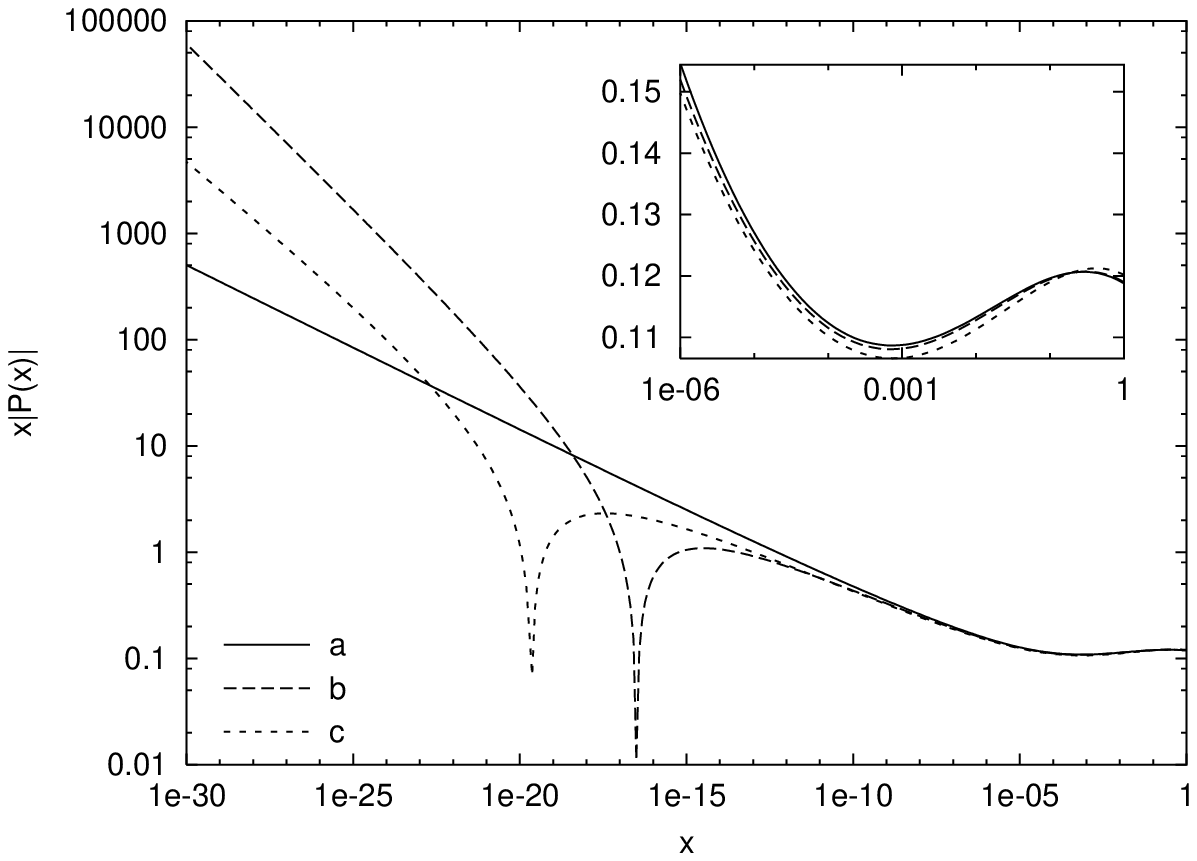, width=\textwidth}
    \caption{Examples of effective splitting functions for $t=9.2$
      (corresponding to $\asb\simeq0.12$)
      and three different combinations of NP parameters: (a) has
      $\tbar=1.0, t_0=2.0$; (b) has $\tbar=1.0, t_0=3$ and (c) has
      $\tbar=2.0, t_0=4.6$. The inset shows the same splitting
      functions on an enlarged scale.}
    \label{fig:astfixed}
}

Figure~\ref{fig:astfixed} is not sufficient to demonstrate that the NP
corrections are truly higher twist. First we consider the `PT part' of
the splitting function.  Figure~\ref{fig:ratioptpower} shows the ratio
of two effective splitting functions, obtained with different
non-perturbative parameter sets (a and b), chosen such that there is
no component with the large NP power growth (which would complicate
the interpretation of the ratio). We see that the NP parameters affect
both the normalisation of the PT splitting function, and the exact
value of the power growth, since the ratio grows as a power. This is
as predicted in the 4-pole model, being due to the shift of the PT
zeroes of the Green's function. We observe that the effect on the
power is relatively small (note the $x$ scale), and that from a
practical point of view it will mostly be the effect on the
normalisation that will be of interest.  The curve at the higher value
of $t$ shows a significantly decreased dependence on the NP
parameters, confirming that their effect is higher-twist.\footnote{The
  attentive reader will have noticed that the modification of the PT
  exponent changes sign at the higher $t$ value --- indeed it turns
  out that the dependence on the NP parameters, while decreasing at
  least as fast as $e^{-t}$, has a non-trivial $t$ dependence for
  moderate $t$ values.}

We also need to demonstrate that the NP power component is higher
twist.  Figure~\ref{fig:astvaried} shows the splitting function for
three different $t$ values, but with the same NP parameters.  The
initial parts of the splitting function now differ since the PT scales
are different; the NP power growth has (approximately) the same power
in the three cases, but its normalisation decreases rapidly with
increasing $t$, roughly as $e^{-t}$, confirming that it too is higher
twist.

\FIGURE{
    \epsfig{file=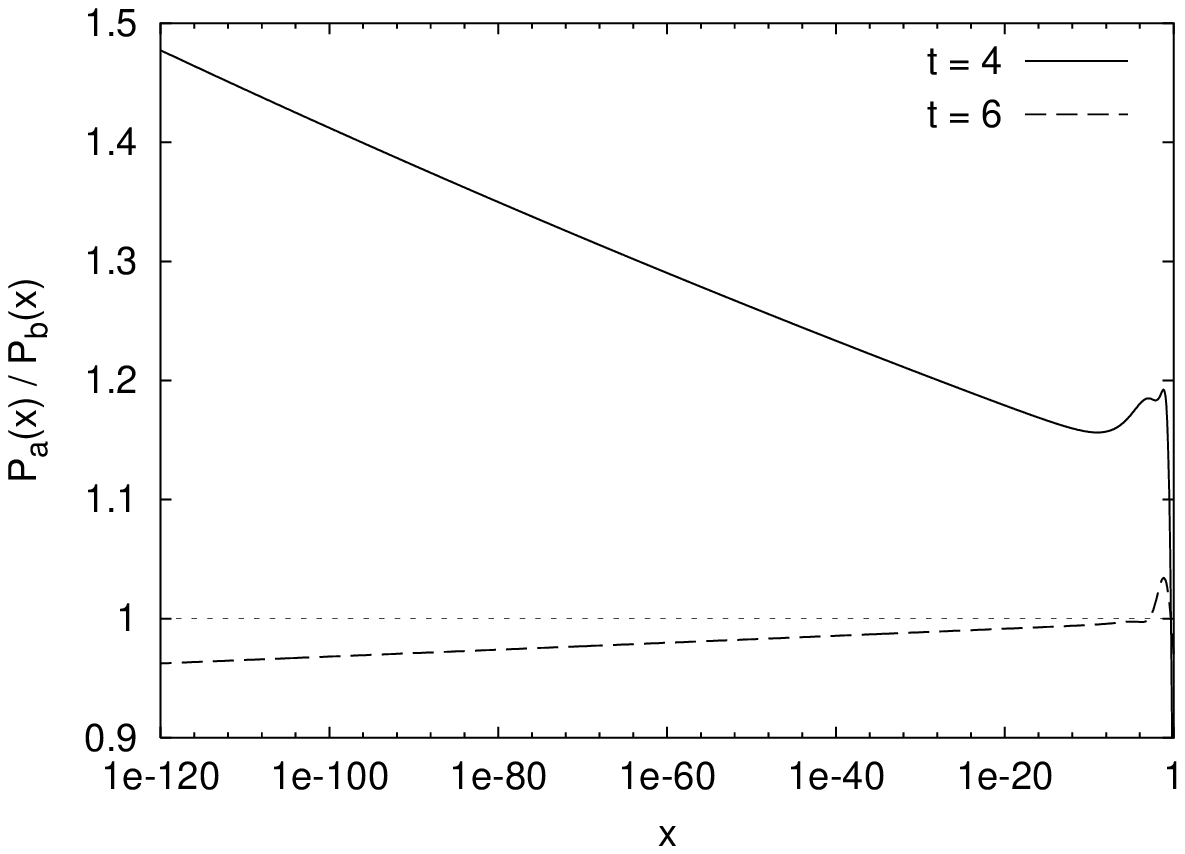}
    \caption{Ratio of splitting functions with the same $t$ but different
      non-perturbative parameters: (a) $\tbar=0.5, t_0=0.8$ and (b)
      $\tbar=1.0, t_0=1.3$.}
    \label{fig:ratioptpower}
}

\FIGURE{
    \epsfig{file=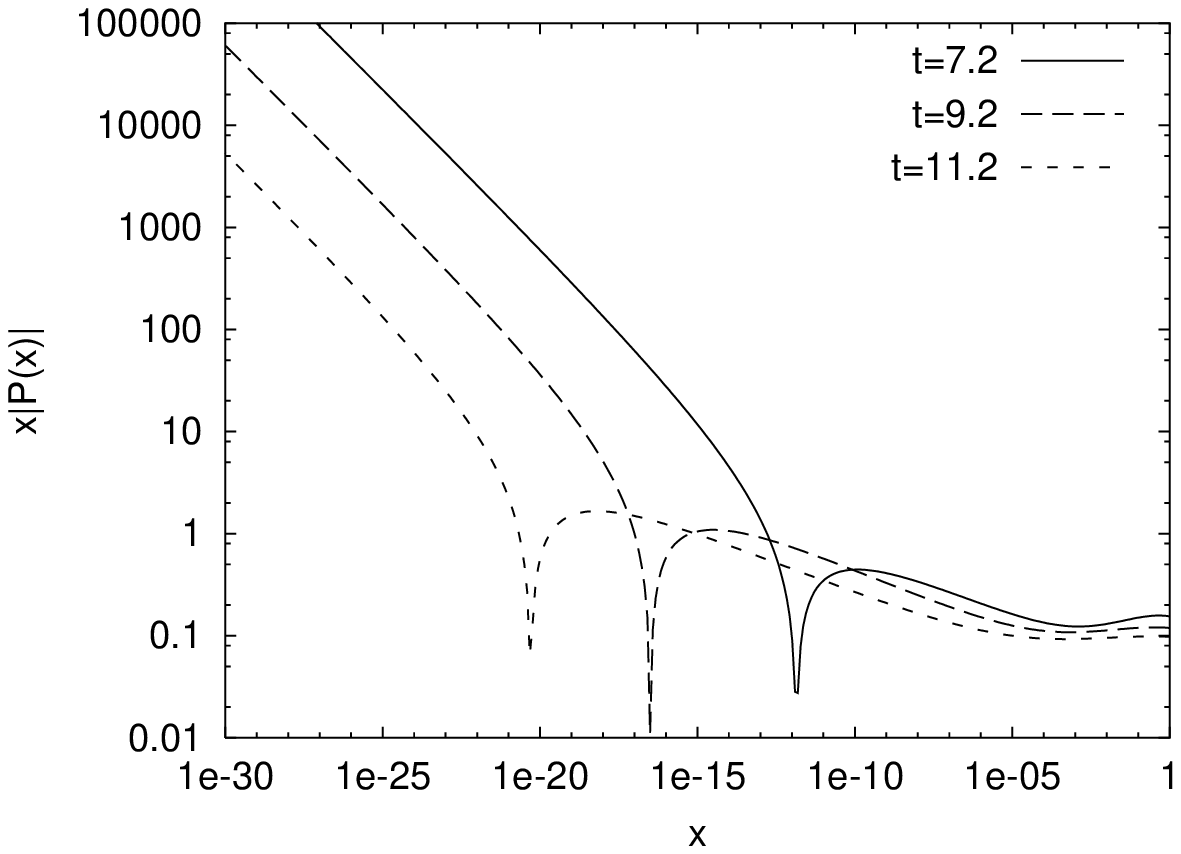}
    \caption{Examples of effective splitting functions for fixed
      non-perturbative parameters ($\tbar=1$, $t_0=3$) and different
      $t$ values, $t=7.2,9.2,11.2$, corresponding to $\asb$ values of
      $0.15$, $0.12$ and $0.097$ respectively}
    \label{fig:astvaried}
}

One of the non-trivial features of the 4-pole model was the
(approximate) equality between the normalisation of the NP power
component, and the quantity $\om_0'-\om_0$, the difference between the
exponent of the power growth and the position of the NP zero
(obtained numerically from the exponent of the NP power growth in the
limit $t\to\infty$). Both quantities are shown in
figure~\ref{fig:deltaom} over a range of $t$ values, illustrating very
clearly their closeness (as well as their higher-twist nature). A
detailed study reveals that the relative difference between them is
consistent with a term of $\cO{\as^2}$, as predicted in
section~\ref{sec:4pole}. We note also that $\om_0$ is found to be
below the NP exponent $\omp$ characteristic of the Green's function,
consistent with it being due to a zero to the left of the leading
singularity.

\FIGURE{
    \epsfig{file=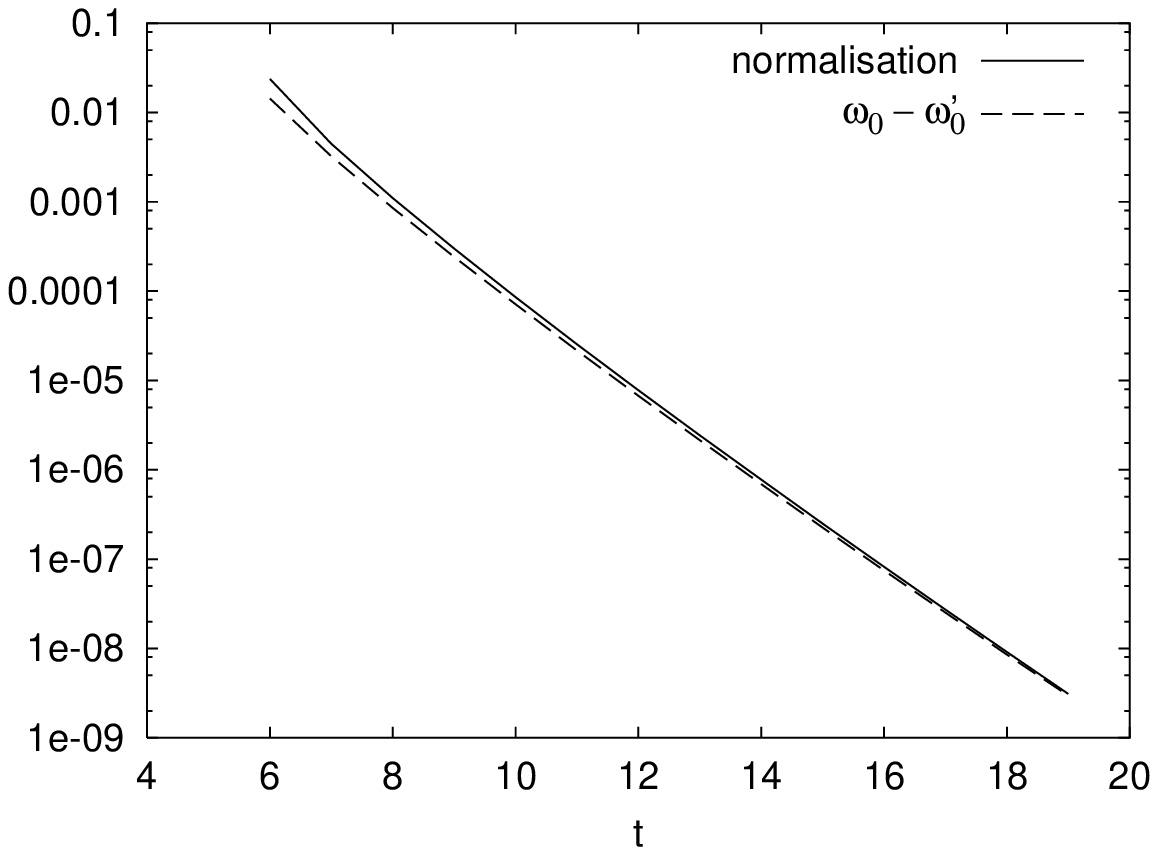}
    \caption{(Minus) the normalisation of the NP power component of the
      splitting function, and the difference $\om_0 - \om_0'$ between
      its exponent and the exponent in the limit $t\to\infty$. Shown
      for the NP parameters $\tbar=2, t_0=5$.}
    \label{fig:deltaom}
}

\section{Conclusions}

For some time now there has been some debate as to whether diffusion
might destroy small-$x$ factorisation \cite{Mueller} or make it
impossible to perturbatively predict the small-$x$ splitting functions
\cite{ABF}. In \cite{CCS1999b} a model was presented which contained
diffusion and the correct collinear limits, but also displayed the
property of exact factorisation. It failed though to make any
statement about (the functional dependence of) the magnitude of any
higher-twist corrections, leaving open the possibility that they could
come to dominate at small $x$, and thus still destroy factorisation.

This paper has presented a much more complete study of the problem,
both by an extension of the model so that it includes leading
higher-twist components, and by the development of numerical
techniques for studying the effective splitting function in the
full (LL with running coupling) BFKL equation.

The basic conclusion of these studies is that higher-twist effects,
while present (and dominant at small $x$ for the effective splitting
function) are truly small \emph{for scaling violations}. An explanation 
of this fact comes from the analysis of the 4-pole collinear model,
whose Green's function is a sum of two terms, each of which with a
NP Pomeron part factorised from the $t$-dependence, the second being
uniformly of higher twist.

Although oversimplified, the collinear example is expected to
exhibit the mechanism at work in a realistic BFKL equation also.
In fact, it can be generalised to the case of $2n$-poles, for arbitrary
$n$, the Green's function being a sum of $n$ terms of higher and
higher twist \cite{cc}. The basic point remains the fact that the
Pomeron singularity, although non-perturbative, is stably factorised
in front of the $t$-dependence in the leading twist term. 



\section*{Acknowledgements}
We wish to thank Martina Taiuti for several discussions in the early
stages of this work.  One of us (GPS) would also like to thank Guido
Altarelli, Yuri Dokshitzer and Al Mueller for interesting
conversations.

\appendix

\section{Green's function of the 4-pole model}

Starting from eq.~\eqref{eq:chi4pole} there are various ways of
deriving the differential equation of the model which differ by the
treatment of the higher twist terms. 

The simplest way is to insert the expression
\begin{equation}
  b \om t - \chi(\ga)
\end{equation}
into the $\ga$-representation of the (regular) solution
\begin{equation}
  \cF_\om(t) = e^{-\frac12 t} F_\om(t) = \int \frac{d\ga}{2\pi i}
  \exp\left[ \left(\ga-\half\right) t - \frac1{b\om}\int^\ga
    \chi(\ga')\,d\ga' \right]
\end{equation}
and to notice that the result vanishes by partial integrations. The
corresponding differential equation is thus obtained by the
replacement $\ga-\half \to \partial_t$. By thus using the identity
\begin{equation}
  \chi(\half + \partial_t) + \frac43 = -\frac{4\partial_t^2 -
    3}{\partial_t^4 - \frac52 \partial_t^2 + \frac9{16}} = -
  \frac{\cmN_t}{\cmD_t} 
\end{equation}
and shifting the $t$ variable to incorporate the constant term
$4/3$, we obtain the Green's function equation
\begin{align}
  \label{eq:4polediff}
  \left[ \om \cmD_t + \cmN_t \asb(t) \right]\, g(t,t_0) &= 
  \left[ \om \cmD_{t_0} + \cmN_{t_0} \asb(t_0) \right]\, g(t,t_0)
  \nonumber \\
  &= -\frac1{\om} \,\cmN_t \, \delta(t-t_0)
\end{align}
where $G_\om(t,t_0) = \om^{-1} \delta (t-t_0) + 
e^{\frac12(t-t_0)} g_\om(t,t_0) \asb(t_0)$ is the gluon
Green's function discussed in the main text.

Note first that $g(t,t_0)$ satisfies a fourth order differential
equation in both the $t$ and $t_0$ variables, for any low-$t$
regularisation of $\asb(t)$, starting from the $1/bt$ expression of
asymptotic freedom (the regularisation depends on the $\tbar$
parameter of the main text). Some care is needed in order to treat the
boundary conditions that $g$ has to satisfy due to the peculiar
distribution occurring in the right-hand side of
eq.~\eqref{eq:4polediff}. This distribution can be taken into account
by assuming $g(t,t_0)$ to be a linear combination of left (right)
regular solutions of the homogeneous equation, as in
\eqref{eq:4poleG}, satisfying at $t=t_0$ the discontinuity
requirements
\begin{equation}
  \label{eq:discont}
  \Delta g = \Delta g'' = 0\,,\quad \Delta g' = -\frac4{\om^2}\,,\quad
  \Delta g'''=-\frac7{\om^2} + \frac{16\asb(t_0)}{\om^3}\,,\qquad (t =
  t_0)\,.
\end{equation}
Given an arbitrary basis $\cmF_L^0(t_0), \cmF_L^1(t_0)$ of
left-regular solutions, the Green's function takes the form
\begin{multline}
  \label{eq:green4}
  g_\om(t,t_0) = \left(  \cmF_R^0(t) \cmF_L^a(t_0) 
                       + \cmF_R^1(t) \cmF_L^b(t_0) ) \right) \Theta(t-t_0)
             \\+   \left(  \cmF_L^0(t) \cmF_R^a(t_0) 
                       + \cmF_L^1(t) \cmF_R^b(t_0) ) \right) \Theta(t_0-t)
\end{multline}
and should also be symmetrical under $t,t_0$ interchange. Because of 
\eqref{eq:green4}, the discontinuity conditions \eqref{eq:discont} at
$t=t_0$ can be viewed as a system of four linear equations in the four
unknowns $\cmF_L^a(t_0)$,  $\cmF_L^b(t_0)$, $\cmF_R^a(t_0)$,
$\cmF_R^b(t_0)$, which can be solved by standard methods.

As a consequence, the left-regular solutions $\cmF_L^a$,  $\cmF_L^b$
occurring in \eqref{eq:4poleG} and satisfying \eqref{eq:discont} are
given by the following expressions
\begin{subequations}
\begin{align}
  \cmF_L^a(t) &= \frac{\det W(\Delta(t),\, \cmF_R^1(t),\, \cmF_L^0(t),\,
    \cmF_L^1(t))}{ \det W(\cmF_R^0 ,\, \cmF_R^1,\, \cmF_L^0,\,
    \cmF_L^1)}\,,
  \\
  \cmF_L^b(t) &= \frac{\det W(\cmF_R^0(t),\, \Delta(t),\, \cmF_L^0(t),\,
    \cmF_L^1(t))}{ \det W(\cmF_R^0 ,\, \cmF_R^1,\, \cmF_L^0,\,
    \cmF_L^1)}\,.
\end{align}
\end{subequations}
Here the $W$'s are the Wronskian matrices of the corresponding
functions, where in the numerators the column vector of derivatives is
replaced, in the proper place, by the discontinuity vector
($i=0,1,2,3$)
\begin{equation}
  \Delta(t) = ( \Delta g^{(i)} ) = -\frac1{\om^2} 
        \left(0,4,0,7-\frac{16\asb(t)}{\om} \right)\,.
\end{equation}
While the Wronskian in the denominator is constant, one can check that
the numerators are indeed solutions of the basic homogeneous equation
in \eqref{eq:4polediff}. Furthermore the expressions \eqref{eq:green4}
turn out to be independent of the choice of the left-regular basis, by
the linearity properties of Wronskian matrices. Finally the Green's
function is determined by symmetry for $t < t_0$. 

The explicit determination of $\cmF_L^a(t_0)$ and $\cmF_L^b(t_0)$, and
of the corresponding Pomeron singularity requires the solution of a
matching problem for scattering in a fourth-order framework,
depending on the regularisation of $\asb(t)$ to the left. The outcome
of this procedure is the separation of the left-regular solutions into
irregular and regular ones on the right, as follows
\begin{subequations}
\begin{align}
  \cmF^0_L(t_0) &= \cmI_R^0(t_0) + \sigma_{00}(\om) \cmF_R^0(t_0) + \cdots\,,\\
  \cmF^1_L(t_0) &= \cmI_R^1(t_0) + \sigma_{10}(\om) \cmF_R^0(t_0) + \cdots\,,
\end{align}
\end{subequations}
where $ \sigma_{00}(\om)$, $ \sigma_{10}(\om)$, \ldots are scattering
coefficients which carry the (non-perturbative) Pomeron singularity
and $\cmI_R^0 \sim (\cmF_R^0)^{-1}$, $\cmI_R^1 \sim (\cmF_R^1)^{-1}$
are the irregular solutions on the right. The relation of the
non-perturbative to perturbative contributions in each of the terms of
eq.~\eqref{eq:4poleG} is thus very similar to that found in the
two-pole model \cite{CCS1999b}.


\end{document}